\documentclass[12pt]{article} 

\usepackage{ol2}
\usepackage[draft]{hyperref}
\usepackage{amsmath}
\usepackage{ulem}
\usepackage{graphicx}

\begin{document}


\title{Characterization of the absolute frequency stability of an individual reference cavity}


\author{T. Liu, Y. N. Zhao, V. Elman, A. Stejskal, and L. J. Wang$^{*}$}

\address{
Institute of Optics, Information and Photonics, Max Planck Research Group, University of Erlangen-Nuremberg, G\"{u}nther-Scharowsky-Str. 1, Bau 24, 91058, Erlangen, Germany \\

$^*$Corresponding author: lwan@optik.uni-erlangen.de}

\begin{abstract}
We demonstrated for the first time the characterization of absolute frequency stability of three reference cavities by cross beating three laser beams which are independently locked to these reference cavities. This method shows the individual feature of each reference cavity, while conventional beatnote measurement between two cavities can only provide an upper bound. This method allows for numerous applications such as optimizing the performance of the reference cavity for optical clockwork.
\end{abstract}



\noindent
 

Frequency stabilized narrow-linewidth lasers are essential as local oscillators in developing optical frequency standards, where their short-term stability makes it possible to resolve the narrow atomic transition\cite{NPLScience,TokyoNature}. Furthermore, they have important applications in high-resolution spectroscopy\cite{NISTPRL05}, tests of fundamental physics\cite{PTBPRL04} as well as interferometric measurements like future space missions including the laser interferometer space antenna (LISA)\cite{LISA04}.

A passive optical cavity serves as the heart of such a narrow-linewidth laser system\cite{DreverAPB83}. Basically, a laser frequency is actively controlled to match a cavity resonance mode which is determined by the effective cavity length. Therefore, the frequency stability of the laser is limited mainly by the geometric change of the cavity length, which is induced technically by the thermal, acoustical, mechanical and seismic noise. The fundamental limit is due to the Brownian motion of cavity mirrors\cite{CampPRL04}. To improve the frequency stability, several innovative designs have been developed to reduce the coupling of the reference cavity to the environmental disturbance\cite{NISTPRL99,HallOL05,PTBAPB06}. 

A conclusive method needs to be applied to evaluate the performance of various reference cavity designs\cite{NPLPRA08,MPQPRA08}. All existing experimental reports on ultra-stable laser compare two cavities to estimate their combined frequency stability. By assuming that these two cavities are identical and indepentent, the performance of an individual cavity can be estimated.

In this Letter, we report for the first time the characterization of the frequency stability of three reference cavities by simultaneously cross-beating three laser beams which are frequency stabilized to respective reference cavities. This experiment follows the idea of ``three-cornered-hat'' method which is widely used in the evaluation of microwave frequency standards\cite{AllanIEEE}. A fractional frequency stability of 6$\times10^{-16}$ with one second interrogation time is observed for a newly developed reference cavity. Furthermore, we investigate the correlation between each of the reference cavities at different time scales. 

\begin{figure}[h]
  \centerline{
    \includegraphics[width = 8.0cm]{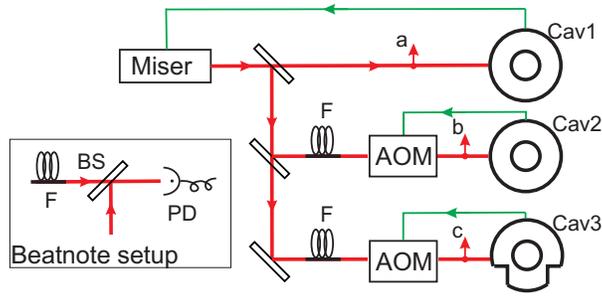}
	}
	\caption{(Color online) Schematic experimental setup for three-cornered-hat measurement. F: single mode fiber; AOM: acousto-optical modulator; PD: photodiode. The red lines indicate the optical paths; the green lines indicate the electronic signal paths. The light beams are extracted from a, b, c points and sent to the respective beatnote setups as shown in the inset at the left side.}
	\label{fig:beatnotesetup}
			  \vspace{-3mm}
\end{figure}

As shown in Fig.\ref{fig:beatnotesetup}, three laser beams are split from one laser source and frequency locked, simultaneously to three reference cavities labeled Cav1, Cav2, and Cav3. In the experiment, Cav1 and Cav2 are the cavities with simple cylindrical bodies but different geometries to avoid common mode noise. According to previous measurements, the short-term stability of such cavities is limited to the 2$\times10^{-14}$ level, which result in a moderate resolution of the narrow-linewidth atomic transition\cite{TaoAPB}. Using the principles developed by the PTB group\cite{PTBAPB06}, we design a new type of optical cavity (Cav3) with a cutoff to reduce the sensitivity to vertical and horizontal mechanical noise. Each cavity is produced of a low thermal expansion material (Cav1 and Cav3: ULE, Cav2: Zerodur) and placed inside a vacuum chamber with $~$10$^{-8}$mbar pressure. Active control system is applied on the shielding box of each cavity to reduce the temperature fluctuation($<$10mK on the vacuum chamber body). Each of the cavities and corresponding optical setup is mechanically isolated by an active vibration isolation platform (AVI). Single-mode optical fiber is used for transferring the beam to neighboring setup. The cutoff frequency of the AVI system is measured to be below 2Hz. This provides excellent mechanical isolation between the cavities and the optical table, which is also helpful to avoid the correlation between different cavities by common mechanical coupling.


A monolithic isolated single-mode end-pumped ring Nd:YAG laser (Miser) is used as the laser source and three beams are derived and locked to the respective cavities. The conventional Pound-Drever-Hall locking technique is applied to realize frequency stabilization. In the case of Cav1, the feedback signal is applied to a Piezo mounted on the top of the laser crystal to control its frequency. With a 50kHz loop bandwidth, the laser frequency is locked to the reference cavity. With a homemade thermal stablization loop, such a stabilized laser system can work continuously over several days. The remaining two laser beams are transfered via a short fiber to the neighboring reference cavities, i.e. Cav2 and Cav3. Each laser beam passes through an optical setup similar to that of Cav1, while the feedback signal is applied to an AOM as the frequency control.

In a typical beatnote setup as in the inset of Fig.1, a locally stabilized beam is beating with a remote beam transferred via a fiber. Due to unexpected coupling to the noisy environment, the fiber could induce a phase noise on the laser mode, causing linewidth broadening. We measure the typical phase noise induced by a 5m fiber used in this experiment by comparing the transmission beam with a laser beam before the fiber. Under a quiet lab condition, a linewidth broadening $<$200mHz was observed. This result suggests that the fiber induced linewidth broadening is not the dominating noise source in the present measurement and can be ignored in later experiments.
The beatnote signal between two lasers is recorded either by a commercial frequency counter(Agilent Inc., 53132A) for the gate time $>$1s or by an analog-digital-converter(Gage Inc, CompuScope 12400) (ADC) for the gate time $<$1s. The sampling rate of the ADC is set to 25MHz/s and the beatnote frequency is down-converted to $~$10kHz to decrease systematic error. Three beat signals are measured with a gate time of $\tau$. This is repeated n times in the experiment, and we can obtain three series of beat frequencies between each two lasers as follows:
\begin{eqnarray}
\nonumber
{y^{ij}_{1}, y^{ij}_{2}, y^{ij}_{3},..., y^{ij}_{n}}\\
{y^{ik}_{1}, y^{ik}_{2}, y^{ik}_{3},..., y^{ik}_{n}}\\	
\nonumber
{y^{jk}_{1}, y^{jk}_{2}, y^{jk}_{3},..., y^{jk}_{n}}
\end{eqnarray}
Here $y^{ij}_{n}$ is the $\textit{n}$-th measured frequency value between cavity $\textit{i}$ and $\textit{j}$.

\begin{figure}[h]
  \vspace{-5mm}
  \centerline{
    \includegraphics[width = 8cm]{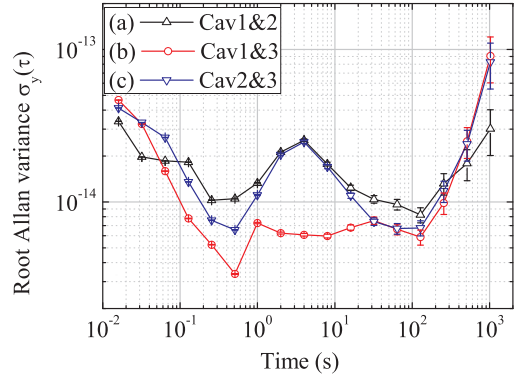}
	}
	  \vspace{-3mm}
	  \caption{(Color online) Relative root Allan variance of the cross-beating frequency between three reference cavities. The error bars in the plot represent the standard deviations concerning the statistical error.}
	\label{fig:TCH-mix}
		  \vspace{-1mm}
\end{figure}

The root Allan variance of each dataset is calculated and the results are shown in Fig.\ref{fig:TCH-mix}. The linear frequency drift is removed. Here we note that this deviation still shows the relative frequency fluctuation between two cavities. 
The beatnote result between Cav1 and 3 shows a better short-term stability than the beatnote results from Cav1$\&$2 and Cav2$\&$3. This might be a hint that the new cavity (Cav3) has a better mechanical insensitivity. However, the result of Cav2$\&$3 let us hardly decide if the new cavity has any improvement. Considering all three datasets, we could speculate on the common features that are originating from the common cavity. For example, we can conclude that a slow vibration at 4s is coming from Cav2 due to a residual vibration of its thermal control loop. But the key point here is that we can not characterize the property of each cavity precisely. And the curve (a) complies with our previous measurement\cite{TaoAPB}.

For convenience, we assume that there is no correlation between all three cavities, then one can estimate the frequency stability of each cavity as follows:
\begin{align}
\sigma^{2}_{i}(\tau)=\frac{1}{2}(\sigma^{2}_{ij}(\tau)+\sigma^{2}_{ik}(\tau)-\sigma^{2}_{jk}(\tau))
\label{eq:TCH}
\end{align}
Here, the subscripts $\textit{i}$, $\textit{j}$ and $\textit{k}$ refer to the three cavities. $\sigma_{ij}(\tau)$ is the relative frequency stability between the cavity $\textit{i}$ and $\textit{j}$, and $\sigma_{i}(\tau)$ is the absolute frequency stability of the particular cavity $\textit{i}$. $\tau$ is the averaging time.

\begin{figure}[h]
  \vspace{-5mm}
  \centerline{
    \includegraphics[width = 8.0cm]{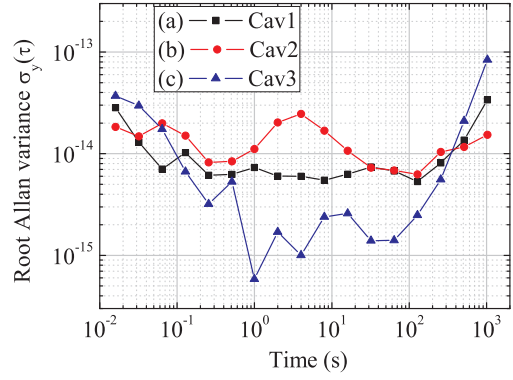}
	}
	  \vspace{-1mm}
	  	\caption{(Color online) Plots of root Allan variances of the individual reference cavities from three-cornered hat measurement. (a) Cavity 1; (b) Cavity 2; (c) Cavity 3.}
	\label{fig:3corneredhat}
		  \vspace{-1mm}
\end{figure}

The absolute frequency stability of the individual cavity is derived from the root Allan variance using Eq. \ref{eq:TCH} and is shown in Fig.\ref{fig:3corneredhat}. It is clear that Cav3 has the best short-term stability in the 1s-100s region, due to the optimized design of the cavity body. The minimum fractional frequency instability of 6$\times$10$^{-16}$ is achieved with an averaging time of one second. For the time $>$100s, the Allan variance rises quickly. This is mainly due to the similar temperature stabilization and aging effect of the new ULE glass. 

Eq. \ref{eq:TCH} is only true when no correlation between the cavities is present. In the opposite case, a covariance term should be included in the right-hand side of the equation. As a result, the conventional two-sample Allan variance will perform the wrong estimation of the frequency stability if a correlation exists between two cavities under investigation. Beside that, other previous works lacked a systematic method to study the correlation between two cavities. Here we take the chance to study the possible correlation between our cavities in different time scales.

To investigate the correlation, we follow the approach in Ref.\cite{TCH}. We choose one of the cavities, for instance cavity $\textit{i}$, as the reference and compare it with the other two at $\textit{M}$ different time instants. The measurement (co)variances are denoted by $S_{jk}$ and $Var_{ij}$, and defined in analogy to Ref.\cite{TCH}, as
\begin{equation}
S_{jk}=\left\langle(y^{ij}-\bar{y}^{ij})(y^{ik}-\bar{y}^{ik})\right\rangle
\end{equation}
The variance $Var_{ij}$ is the covariance $S_{jj}$ measured with the cavity $\textit{i}$ as the reference.

The uncorrelation condition implies that:
\begin{equation}
0<S_{jk}<<Var_{ij}, 0<S_{jk}<<Var_{ik}
\label{eq:criteria}
\end{equation}
The (co)variances of three datasets are calculated and they show very consistent results at different time scales. For convenience, part of the results are listed in Table \ref{tab:Correlation}. 

\begin{table}
		  \vspace{-5mm}
	\centering
	\caption{Correlation\protect\label{tab:Correlation}}
	\vspace{2mm}
			\begin{tabular}{rcccccc}
			\hline
time&$S_{23}$&$S_{12}$&$S_{13}$&$Var_{12}$&$Var_{13}$&$Var_{23}$\\
\hline
1s&-10.09&6.19&25.81&106.71&31.93&73.90\\
16s&-23.05&3.81&64.80&92.12&27.75&72.98\\
128s&-15.57&5.57&22.95&40.79&20.84&27.41\\
\hline
		\end{tabular}
\end{table}

From the table, we can see the covariance $S_{23}$ is less than zero at all three time scales. This indicates that there is always an anonymous coupling between Cav2 and Cav3. The reason for this coupling is still under investigation. Then Eq. \ref{eq:criteria} is inspected for the rest cavities. For example, with one second instant, the covariance $S_{12}$=6.19 is less than the variance $Var_{13}$=31.93 and $Var_{23}$=73.90. So there is no correlation between Cav1 and Cav2. After inspecting all of the cases, we conclude that a correlation exists only between Cav2 and Cav3. Beside this, the absolute frequency stability of the reference cavity can be calculated more precisely by considering the correlation terms.

In conclusion, we have demonstrated for the first time the characterization of the absolute frequency stability of three laser beams independently locked to respective reference cavities. A fractional frequency stability of 6$\times10^{-16}$ with one second averaging time is observed for a new reference cavity. The correlations between each of the reference cavities at different time scales are investigated as well. The method shown in this Letter can be applied to launch numerous applications such as optimizing the performance of the reference cavity for optical clockwork. 

We thank Z. H. Lu and H. Schwefel for helpful discussions.

%

%
%

\newpage


\begin{list}{}{}

\item Fig. 1. (Color online) Schematic experimental setup for three-cornered-hat measurement. F: single mode fiber; AOM: acousto-optical modulator; PD: photodiode. The red lines indicate the optical paths; the green lines indicate the electronic signal paths. The light beams are extracted from a, b, c points and sent to the respective beatnote setups as shown in the inset at the left side.

\item Fig. 2. (Color online) Relative root Allan variance of the cross-beating frequency between three reference cavities. The error bars in the plot represent the standard deviations concerning the statistical error.

\item Fig. 3. Plots of root Allan variances of the individual reference cavities from three-cornered hat measurement. (a) Cavity 1; (b) Cavity 2; (c) Cavity 3.

\end{list}
\newpage

\newpage

\newpage

\newpage


%

%

\end{document}